\documentclass[aps,twocolumn]{revtex4}
\usepackage{graphicx}
\usepackage{amsmath}
\usepackage{amssymb}
\newcommand{\be}{\begin{equation}}
\newcommand{\ee}{\end{equation}}
\newcommand{\bea}{\begin{eqnarray}}
\newcommand{\eea}{\end{eqnarray}}
\newcommand{\pr}{\partial}
\newcommand{\nno}{\nonumber}
\newcommand{\bse}{\begin{subequations}}
\newcommand{\ese}{\end{subequations}}
\newcommand{\cphi}{\phi_{cl}}
\begin{document}
\title{Stability analysis of the Randall-Sundrum braneworld in presence of bulk scalar}
\author{Debaprasad Maity\footnote{E-mail: tpdm@mahendra@iacs.res.in}, Soumitra SenGupta
\footnote{E-mail: tpssg@mahendra@iacs.res.in}}
\address{Department of Theoretical Physics, Indian Association for the
Cultivation of Science,\\
Calcutta - 700 032, India}
\author{Sourav Sur\footnote{E-mail: sourav@iopb.res.in}}
\address{Institute of Physics, Bhubaneswar - 751 005, India}
\begin{abstract}
The stability problem of Randall-Sundrum braneworld is readdressed in the light of 
stabilizing bulk scalar fields. It is shown that in such scenario the instability 
persists because of back-reaction even when an arbitrary potential is introduced 
for a canonical scalar field in the bulk. It is further shown that a bulk scalar 
field can indeed stabilize the braneworld when it has a tachyon-like
action. The full back-reacted metric in such model is derived and a proper 
resolution of the hierarchy problem (for which the Randall Sundrum scenario was 
originally proposed) is found to exist by suitable adjustments of 
the parameters of the scalar potential.
\end{abstract}
\maketitle

Standard model for strong, weak and electromagnetic interactions based on 
the gauge group $ SU(3)\times SU(2)\times U(1)$ has been extremely 
successful in explaining physical phenomena upto TeV scale. Such a model,
however, encounters the well known fine tuning problem in connection with 
Higgs mass related to the gauge hierarchy problem which
refers to the vast disparity between the weak and Planck scale. 
By invoking supersymmetry one can resolve this problem at the expense of 
incorporating a large number of (hitherto unseen) superpartners in the theory. 
In an alternative approach, theories with extra spatial dimension(s) have 
attracted a lot of attention because of the new geometric approach to solve 
the same problem and also the distinct features from ordinary Kaluza-Klien theory.
In such models the standard model fields are localized on a 3+1 dimensional brane 
\cite{arkani,antoniadis,witten,rs,lykken,cohen,kaloper} where gravity can 
propagate in the bulk spacetime. One of the most theoretically appealing and 
phenomenologically interesting model in this context was proposed by Randall and 
Sundrum (RS) \cite{rs} where the mass hierarchy emerged naturally in an exponentially 
warped geometry along the extra dimension. The model contains two 3+1 dimensional 
branes sitting at the two orbifold fixed points where the single extra dimension 
in a 4+1 dimensional bulk has been compactified on a $S_1/Z_2$ orbifold. The model 
contains two parameters namely the bulk cosmological constant $\Lambda$ and the 
brane separation $r_c$. The geometry and the Higgs mass is warped exponentially by 
a dimensionless parameter $k r_c$, where $k = \sqrt{-\Lambda/24M^3}$, $M$ being 
the 5D Planck mass. For the desired warping one should have $k r_c \sim 11$. If 
$k \sim$ Planck mass then, $r_c$ should have a stable value near Planck length.
To stabilize the value of $r_c$, Goldberger and Wise (GW) \cite{gw} proposed a 
simple mechanism by introducing a minimally coupled massive scalar field in the bulk.
Later several other works have been done in this direction 
\cite{luty-sundrum,maru,ssg,kogan}. 
In the original work of GW the effect of back-reaction of the scalar field on the 
background metric was neglected. Such back-reaction was later included in 
subsequent works and a modified solution for the metric was found \cite{gub,csaki}. 
In this paper we carefully re-examine the stability issue in the back-reacted RS 
model when the scalar field is introduced in bulk. We show on a very general ground 
that stabilization is not possible with a minimally coupled scalar field in the bulk. 
Following GW we subsequently show that if the dependence of scalar field action on the
extra coordinate is tachyon-like then we can stabilize the brane world and the stabilized 
value of $r_c$ can produce the desired hierarchy from Planck scale to TeV scale. 

Let us first consider the following bulk action \cite{gw,gub,csaki,bulk}
\bea \label{bulkac}
& & S = \int~ d^5 x ~\sqrt{- G} \Big{[} - M^3 R ~+~ \sum_{i = 1}^n \left(\frac 1 2 \pr_A \phi_i
~ \pr^A \phi_i\right) \nno\\ & & - V (\{\phi_i\})\Big{]}  
- \int d^4 x ~dy~ \sqrt{- g_a} \delta (y - y_a) \lambda_a (\{\phi_i\}) 
\eea
where $\{\phi_i\} = \{\phi_1 \cdots \phi_n\}$ are a set of $n$ scalar fields with 
a mixed potential
$V (\{\phi_i\})$ and the index $a$ runs over the brane locations. The 
corresponding brane potentials are denoted by $\lambda_a$.

Taking the line element in the form
\be \label{metans}
ds^2 ~=~ e^{- 2 A(y)} \eta_{\mu \nu} dx^\mu dx^\nu ~-~ dy^2 
\ee
the field equations are given as
\bse \label{feq}
\bea 
\phi_i'' ~-~ 4 A' \phi_i' &=& \frac{\pr V}{\pr \phi_i} ~+~ \sum_a \frac{\pr \lambda_a}
{\pr \phi_i} \delta (y - y_a) \\
4 A'^2 - A'' &=& - \frac V {3 M^3} - \frac 1 {6 M^3} \sum_a \lambda_a
\delta (y - y_a) ~~~~~~~~~~~\\
4 A'^2 ~-~ 4 A'' &=& - \frac V {3 M^3} ~-~ \frac 1 {2 M^3} \sum_i \phi_i'^2  \nno\\ 
&-& \frac 2 {3 M^3} \sum_a \lambda_a \delta (y - y_a) 
\eea
\ese
where prime $\{'\}$ denotes partial differentiation with respect to the extra spatial
coordinate $y$.

The boundary conditions are
\be \label{bc}
\left[\phi_i'\right]_a ~=~ \frac {\pr \lambda_a}{\pr \phi_i}~;~~~
\left[A'\right]_a ~=~ \frac 1 {6 M^3} \lambda_a (\{\phi_i\}) .
\ee
Denoting $C = 1/24 M^3$, it follows
\bse \label{efffeq}
\bea 
A'^2 &=& 2 C \left(\sum_i \phi_i'^2 - 2 V\right) ;~
A'' = 4 C \sum_i \phi_i'^2\\
\phi_i'' &=& 4 A' \phi_i' + \frac{\pr V}{\pr \phi_i} 
\eea
\ese

In order to obtain analytic closed form solutions we resort to the particular
class of effective potentials as in refs. \cite{gub,cvetic}:
\be \label{bulkpot}
V (\{\phi_i\}) ~=~ \frac 1 8 \sum_{i = 1}^n \left[\frac{\pr W}{\pr \phi_i} 
\right]^2 ~-~ 2 C W^2 
\ee
where 
\be
W (\{\phi_i\}) ~=~ \sum_{i = 1}^n W_i (\phi_i)
\ee
is a sum of $n$ superpotentials $\{W_1 (\phi_1), \cdots, W_n (\phi_n)\}$.

Eqs.(\ref{efffeq}) lead to the first order equations:
\be \label{soleq}
\phi_i' ~=~ \frac 1 2 \frac{\pr W}{\pr \phi_i} ~;~~~
A' ~=~ 2 C W .
\ee
Following GW \cite{gw}, we calculate the effective potential $V_{eff}$ on 
the 3-brane as:
\bea \label{effpot}
& & V_{eff} = 2 \int_0^{y_\pi} dy ~ e^{- 4 A(y)} \Big[\frac 1 2 \sum_{i = 1}^n  
\pr_A \phi_i \pr^A \phi_i - V(\{\phi_i\})\Big] \nno\\
& & -~ e^{- 4 A(0)} \lambda_0 (\{\phi_i^0\}) ~-~ e^{- 4 A(y_\pi)} 
\lambda_\pi (\{\phi_i^\pi\})
\eea 
where the Planck brane is located at $y = 0$ and the visible brane
is at $y = y_\pi$, and we have used the notations: $~\phi_i^0 \equiv \phi_i (y = 0) ,
~\phi_i^\pi \equiv \phi_i (y = y_\pi)~$ for $i = 1, \cdots, n$. Using the boundary 
conditions (\ref{bc}) which now take the form
\bse \label{effbc}
\bea
\left[W (\{\phi_i\})\right]_{0} = 2 \lambda_0 
(\{\phi_i\}) &;& \left[\frac{\pr W}{\pr \phi_i}\right]_{0} 
= 2 \frac{\pr \lambda_0}{\pr \phi_i} ~~~~~\\
\left[W (\{\phi_i\})\right]_{y_\pi} = 2 
\lambda_\pi (\{\phi_i\}) &;&
\left[\frac{\pr W}{\pr \phi_i}\right]_{y_\pi} = 2 
\frac{\pr \lambda_{\pi}}{\pr \phi_i} ~~~~~
\eea
\ese
the brane potentials $\lambda_0 (\phi)$ and $\lambda_\pi (\phi)$ are given by \cite{gub}
\bse \label{brpot}
\bea
\lambda_0 (\{\phi_i\}) &=& W (\{\phi_i^0\}) ~+~ 
\sum_{i = 1}^n \left[\frac{\pr W}{\pr \phi_i^0} \left(\phi_i - 
\phi_i^0\right)\right] \nno\\ &+&~ \sum_{i = 1}^n \gamma_{0 i}^2 \left(\phi_i - 
\phi_i^0\right)^2 \\
\lambda_\pi (\{\phi_i\}) &=& - W (\{\phi_i^\pi\}) ~-~ 
\sum_{i = 1}^n \left[\frac{\pr W}{\pr \phi_i^\pi} \left(\phi_i - 
\phi_i^\pi\right)\right]\nno\\ &+& ~ \sum_{i = 1}^n \gamma_{\pi i}^2 \left(\phi_i - \phi_i^\pi\right)^2 
\eea
\ese
where the constants $\gamma_{0 i}$ and $\gamma_{\pi i}$, (for $i = 1, \cdots, n$) are 
parameters of various potentials. 

Now, in order to have extremum for the above effective potential at some value of 
$y_\pi ~$($= y_s$, say) 
 \be \label{1st}
\left[\frac{\pr V_{eff}}{\pr y_\pi}\right]_{y_\pi= y_s} =~ - 4 C 
e^{- 4 A (y_s)} W (y_s)^2 ~=~ 0 .
\ee
which implies $W(y_{\pi}) ~=~ 0$ as well as $\pr^2 V_{eff}/\pr y_\pi^2 ~=~ 0$ at 
$y_{\pi} = y_s$. On the other hand, the third derivative is given by
\be \label{3rd}
\left[\frac{\pr^3 V_{eff}}{\pr y_\pi^3}\right]_{y_\pi= y_s} =~ - 4 C e^{- 4 A (y_s)}
\left[\frac{\pr W (y_\pi)}{\pr y_\pi}\right]_{y_\pi= y_s}^2 
\ee
which is clearly non-vanishing in accord with the above boundary conditions. Therefore,
no extremum exists and we only have a point of inflection at the value of $y_\pi = y_s$ 
for which $W$ vanishes. 

Let us now resort to a general solution ansatz for the metric function $A (y)$ and single 
scalar field $\phi$ so that the standard RS \cite{rs} solution is obtained in the limit 
where the scalar $\phi (y)$ becomes trivial: 
\be \label{gensol}
A (y) ~=~ k y ~+~ f \left(\cphi (y)\right)~;~~ \phi = \cphi(y, a_1, a_2) 
\ee
where $a_1, a_2$ are the initial parameters of the theory and constant $k \sim$ 5D Planck scale as in the RS picture. 

The field equations (\ref{efffeq}) yield
\bse \label{eom}
\bea 
\cphi'^2 &=& \frac 1 {4 C} ~\frac{\pr^2 f}{\pr y^2} \\
V &=&  \frac 1 {8 C} ~\frac{\pr^2 f}{\pr y^2} ~-~ \frac 1 {2 C} \left(k ~+~ \frac{\pr f}{\pr y}
\right)^2 .
\eea
\ese
In the limit $f \rightarrow 0$, $V \rightarrow - 12 M^3 k^2 = \Lambda$ which
gives the standard AdS$_5$ bulk geometry of RS model.

>From Eqs.(\ref{bc}), the boundary values of the brane potentials and their 
first derivatives are obtained as
\bse
\bea
\left[\lambda_0\right]_0 = 2 C \left[k + \frac{\pr f}{\pr y}
\right]_{0} ;~  
\left[\frac{\pr \lambda_0}{\pr y}\right]_{0} = 2 C \left[\frac{\pr^2 f}
{\pr y^2}\right]_{0} ~~~~~~~~~~\\
\left[\lambda_\pi\right]_{y_\pi} = - 2 C \left[k + \frac{\pr f}{\pr y}
\right]_{y_\pi} ;~
\left[\frac{\pr \lambda_\pi}{\pr y}\right]_{y_\pi} = - 2 C \left[\frac{\pr^2 f}
{\pr y^2}\right]_{y_\pi} .
\eea
\ese

At a stable point $y_\pi = y_s$, the effective 4D potential (\ref{effpot}) 
has a vanishing first derivative 
\be
\left[\frac{\pr V_{eff}}{\pr y_\pi}\right]_{y_\pi = y_s} = - \frac{e^{- 4 A (y_\pi)}} C 
\left[k ~+~ \frac{\pr f}{\pr y}\right]^2_{y = y_s} = 0 
\ee
which at once implies that the second derivative $\pr^2 V_{eff}/\pr y_\pi^2$ also vanishes 
at the stable point $y_s$. The third derivative, however, is given at $y_\pi = y_s$ as
\be
\left[\frac{\pr^3 V_{eff}}{\pr y_\pi^3}\right]_{y_\pi = y_s} ~=~ - \frac{2 e^{- 4 A (y_s)}}
C \left[\frac{\pr^2 f}{\pr y^2}\right]^2_{y = y_s}
\ee
which is, of course, non-vanishing in consistence with Eq.(\ref{eom}a).  

Considering, for example
\be
f ~=~ a_1 ~exp \left(- 2 a_2 y\right)
\ee
we find from Eqs.(\ref{eom})
\bea
\phi_{cl} = \sqrt{ a_1} e^{- a_2 y} ~;~~
V = \frac 1 8 \left[ \frac {\pr W}{\pr \phi} \right]^2 - 2 C W^2
\eea
where $W = 12 k M^3 - a_2 \phi^2$.
Clearly, one gets back the same results as discussed in \cite{gub}.

We now consider a tachyon-like scalar in the bulk \cite{sen,sk}. The action is given as 
\bea
S &=& - \int d^5 x ~\sqrt{- G} \left[R + V \sqrt{1 - \pr_{A}\phi \pr^{A} \phi}\right] \nno\\
&-& \int d^4 x ~\sqrt{- g_i} ~\lambda_i(\phi)
\eea
where $ V $ is the potential of the field. 

With a similar metric ansatz the field equations take the form 
\bse \label{teqm}
\bea
&& A'^2 ~=~ -\frac 1 {12} \frac V {\sqrt{1 + \phi'2}} ; ~~ A'' ~=~ \frac 1 6 
\frac { V \phi'^2}{\sqrt{1 + \phi'^2}} ~~~~~~~\\ 
&& \frac {\phi''}{(1 + \phi'^2)}  ~-~ 4 A' \phi' ~-~ \frac 1 V \frac {\pr V}{\pr \phi} ~=~ 
0 ~~~~~~~
\eea
\ese 
and the corresponding boundary conditions are
\bea \label{tbc} 
\left[A'\right]_i ~=~ \frac 1 6 \lambda_i(\phi) ~;~
\left[\frac { V \phi'}{(1 + \phi'^2)^{3/2}}\right]_i =~ 
\frac {\pr \lambda_i(\phi)}{\pr \phi}
\eea
where $\{i\}$ corresponds to the brane locations at orbifold fixed points 
along the extra direction.

Proceeding along the same line as in the previous section, we obtain from 
Eq.(\ref{teqm})
\bea
\phi'^2 ~=~ - \frac {A''}{2 A'^2} ~;~~~~
V ~=~ -12 A'\sqrt{ A'^2 - A''/2} .
\eea
Let us now consider the simple solution ansatz as
\be
A ~=~ k y + a_1 \left(y - a_2\right)^2
\ee
where $ k, a_1, a_2 $ are the parameters of the theory. We get 
\bse
\bea
\cphi'^2 = - \frac {a_1} {\left[y + 2 a_1 (y - a_2)\right]^2} ~~~~~~~~~~~~~~~~~~~~~~~~~~~~~~~~\\
V = - 12 \left[k + 2 a_1 (y - a_2)\right] \sqrt{1 - \frac {a_1} {k + 2 a_1 (y - a_2)}} .~~~
\eea
\ese

These equations imply that $a_1 = 0$ gives $V = -12 k$ and $\phi =$ constant, indicating 
pure RS background. We are, however, interested in $a_1 \ne 0$, in which case a real 
solution for $\phi$ can be obtained only when $a_1$ is negative:
\be 
\phi_{cl} ~=~ \pm \left(- 4 a_1\right)^{- 1/2} ~\ln \left[k + 2 a_1(y - a_2)\right] .
\ee
After a long but straightforward calculation following GW \cite{gw}, we get
\be
\frac{\pr V_{eff}}{\pr y_{\pi}} = - 72 e^{- 4 A} \left[\frac {- A''^2 + 4 A'^4 - A''A'^2}
{12 A'^2 - 6 A''}\right]_{y = y_{\pi}} .
\ee
So the stable point is at
\be
y_s ~=~ \sqrt{ \frac 1 {2 |a_1| (\sqrt{5} +1)}} ~+~ \frac k { 2 |a_1|} ~+~ a_2 .
\ee 
The corresponding metric function, upon setting $|a_1| = m^2$, is given by
\be
A ~=~ \frac {k^2}{4 m^2} ~+~ k a_2 ~-~ 0.154 
\ee
It is now clear that one can easily get the acceptable hierarchy by appropriately
choosing the values of $k/m$ and $k a_2$ with the stabilized radius $y_s$.

Our work clearly reveals an inherent instability in Randall-Sundrum two brane model.
It is shown that just by introducing a canonical scalar field in the bulk, as initially 
suggested by GW, stabilization can not be achieved even with an arbitrary potential if 
full back-reaction of the scalar field on the metric is considered. However, 
we also show that in such models the two-brane separation can indeed be stabilized
if the bulk scalar field has a tachyon-like action with respect to the bulk coordinate. 
We finally show that the full back-reacted metric in such case can also yield the desired 
warping from Planck scale to TeV scale and thus resolves the fine tuning problem of the 
Higgs mass in a stable braneworld scenario.

\vskip .1in
\noindent
{\bf {\large Acknowledgment}}

DM acknowledges Council of Scientific and Industrial Research, Govt. of India for
providing financial support.

\end{document}